\documentclass[12pt, a4paper, oneside]{Thesis} 

\graphicspath{{Pictures/}} 

\usepackage[square, numbers, comma, sort&compress]{natbib} 
\usepackage{graphicx}
\usepackage{caption}

\usepackage[T5]{fontenc}
\usepackage[utf8]{inputenc}
\usepackage{comment}
\usepackage{cite}
\usepackage{hyperref}
\usepackage[hyphenbreaks]{breakurl}
\usepackage{subfigure}
\usepackage{amsmath}
\usepackage{algorithm}
\usepackage{algorithmic}
\hypersetup{urlcolor=black, colorlinks=true} 
\title{\ttitle} 

\begin{document}

\frontmatter 

\setstretch{1.75} 

\fancyhead{} 
\rhead{\thepage} 
\lhead{} 

\pagestyle{fancy} 

\newcommand{\HRule}{\rule{\linewidth}{0.5mm}} 

\hypersetup{pdftitle={\ttitle}}
\hypersetup{pdfsubject=\subjectname}
\hypersetup{pdfauthor=\authornames}
\hypersetup{pdfkeywords=\keywordnames}


\begin{titlepage}
\begin{center}

\textsc{\LARGE \univname}\\[1.5cm] 
\textsc{\Large MASTER'S THESIS}\\[0.5cm] 

\HRule \\[0.4cm] 
{\huge \bfseries \ttitle}\\[0.4cm] 
\HRule \\[1.5cm] 

\begin{minipage}{0.4\textwidth}
\begin{flushleft} 
\emph{By:}
\href{http://www.johnsmith.com}{\authornames} 
\end{flushleft}
\end{minipage}
\begin{minipage}{0.4\textwidth}
\begin{flushright} 
\emph{Supervisor:}
\href{http://www.jamessmith.com}{\supname} 
\end{flushright}
\end{minipage}\\[3cm]

\large \textit{A thesis submitted in partial fulfilment of the requirements \\ for the degree of \degreename}\\[0.3cm] 
\textit{in the}\\[0.4cm]
 \facname\\\deptname\\[2cm] 

{\large June 2014}\\[4cm] 

\vfill
\end{center}

\end{titlepage}

\addtotoc{Abstract} 

\abstract{\addtocontents{toc}{\vspace{1em}} 

Heterogeneous Networks (HetNets) are introduced by the 3GPP as an emerging technology to provide high network coverage and capacity. The HetNets are the combination of multilayer networks such as macrocell, small cell (picocell and femtocell) networks. In such networks, users may suffer significant cross-layer interference. To manage the interference the 3GPP has introduced Enhanced Inter-Cell Interference Coordination (eICIC) techniques, Almost Blank SubFrame (ABSF) is one of the time-domain technique in the eICIC solutions. We propose a dynamically optimal ABSF framework to enhance the small cell user downlink performance while maintains the macro user downlink performance. We also study the mechanism to help the small cell base stations cooperate efficiently in order to reduce the mutual interference. Via simulation, our proposed scheme achieves a significant performance and outperforms the existing ABSF frameworks.

}

\clearpage 


\setstretch{1.75} 

\acknowledgements{\addtocontents{toc}{\vspace{1em}} 

A lot of people have in different ways been helping and supporting me throughout this research work.

First of all, I would like to express my special appreciation and thanks to my advisor, Prof. Sungoh Kwon, who had provided excellent guidance and given me full freedom to work in the field I was interested in. He has been an excellent mentor and very supportive throughout. Without his critical comments and suggestions, none of my work would have been possible.

I would also like to thank my committee members, Prof. Chong-Koo An, Prof. Sungoh Kwon, and Prof. Sunghwan Kim for offering me their creative and valuable comments and serving as my committee members. I would like to acknowledge colleague Quoc Khanh Dang for helpful discussions about the Kalman Filter theory and Matlab simulation toolbox on my research work.

I am also thankful to my brothers and friends Quoc Khanh Dang, Anh Tuan Duong, Quoc Hoan Tran, Minh Luan, Van Duc Le, Ngoc Bach Hoang, Ngoc Hoan Le, Duong Toan for memorable moments and helping me stay sane through these difficult time during my stay at University of Ulsan, and many others from Vietnam: Anh Tuan Vu, Thuc Van Do, Hai Binh for their understand and support.

Last but not at least, I am very grateful for the endless love, support, and encouragement from my whole family. I cannot adequately express the gratitude I feel towards my father who had a great desire for my higher studies, my mother for her endless love, my young sister for supporting me in all regards.

\hfill \emph{Đi khắp thế gian không ai tốt bằng mẹ,}

\hfill \emph{Gánh nặng cuộc đời không ai khổ bằng cha.}

\hfill Ulsan, June 2014.

}
\clearpage 


\pagestyle{fancy} 

\lhead{\emph{Contents}} 
\tableofcontents 

\lhead{\emph{List of Figures}} 
\listoffigures 

\mainmatter 

\pagestyle{fancy} 



\chapter{Introduction to Heterogeneous Networks} 

\label{Chapter1} 

\lhead{Chapter 1. \emph{Introduction to Heterogeneous Networks}} 


\section{Introduction}
\label{Intro}
\begin{figure}
  \centering
  \includegraphics[scale=0.5]{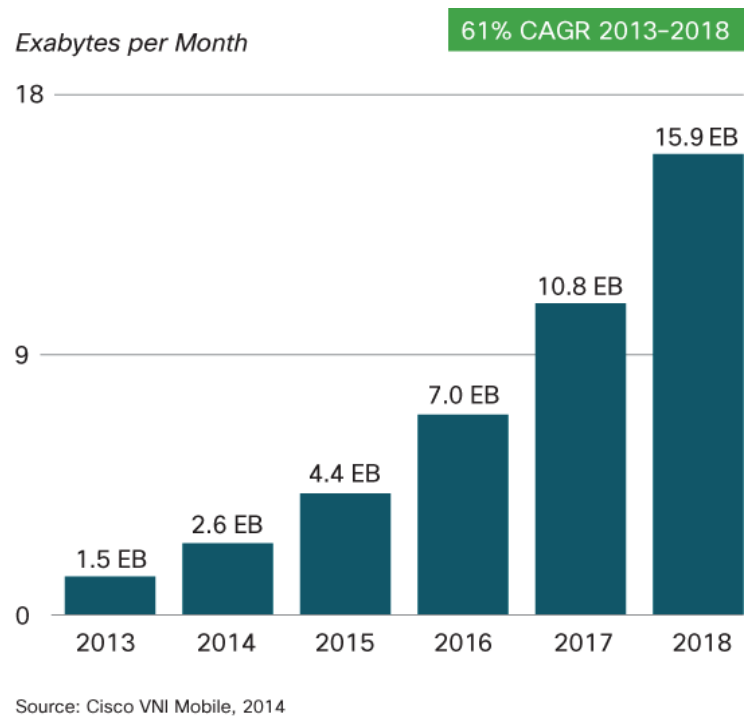}\\
  \caption{Cisco Forecasts 15.9 Exabyte per month of Mobile Data traffic by 2018~\cite{Cisco2013}}\label{lb-MobileData}
\end{figure}

As increasing the number of mobile devices such as smart phones, tablets, and other media devices, the demand for the massive data traffic has been increased. Overall mobile traffic is forecasted to reach 15.9 exabyte per month by 2018, approximately 11 times as much as it was in 2013, according to “Cisco Visual Networking Index: Global Mobile Data Traffic Forecast Update, 2013-2018”~\cite{Cisco2013}. Mobile data traffic will increased at a compound annual growth rate (CAGR) of 61 percent from 2013 to 2018 as shown in Fig.~\ref{lb-MobileData}. In order to meet the demand for data traffic, Third Generation Partnership Project (3GPP) Long Term Evolution-Advanced (LTE-A) has introduced the Heterogeneous Network (HetNet)~\cite{Khandekar2012, TranSS2012}.

LTE based on Orthogonal Frequency Division Multiple Access (OFDMA) is introduced to provide high spectrum efficiency, low latency, and high peak data rates in 2009 as part of the 3GPP Release 8 and 9 specifications~\cite{sesia2009}. LTE was extended to improve the network performance and capacity by introducing the LTE-A in Release 10 and 11 specifications~\cite{ghosh2010S, Hu2012, Khandekar2012}. LTE-A includes uplink and downlink multi-antena (MIMO) technologes, coordinated multi-cell transmission and reception (CoMP), bandwidth extension with carrier aggregation (CA), relay nodes (RN) and heterogeneous networks (HetNets). The MIMO, CoMP, and CA techniques can help improve network performance in some ways, but not offer significant enhancements. HetNet is a key solution in LTE-A to enable the network to reach the theory capacity limits by increasing the quality of the link caused by small distance between users and serving base station.

Currently, the wireless cellular networks consist of only macro base station to serve all users called homogenous networks. In such networks, all macro base stations have same characteristics such as transmission power, antenna patterns, noise parameters, propagation model, and connect each other through similar backhaul connectivity. The users under the coverage of serving macro base station may experience interference from neighbor macro base station. Hence, the location of macro base station are carefully studied before real deployment, and each macro base stations are optimally configured to maximize the network coverage and limit the interference between each other. Due to the requirement of high data traffic, the number of macro base station to be deployed becomes larger that leads to the cost of network services and the difficulty of deployment in some specific geography. Hence, HetNets were introduced as an flexible, low-cost, efficient solution to increase the area network capacity and coverage.

HetNet consists of multiple types of access nodes in wireless network such as macro cell and small cells (i.e micro cells, pico cells, and femto cells). Small cells are deployed underlaid of macro cell in order to improve spectral efficiency per unit area and per link. The macrocell is responsible for providing the overall cellular network coverage. The macro base station is expected as a largest base station with maximum transmit power of 46 dBm, called Macro evolved NodeB (MeNB) in LTE/LTE-A system. The microcells are known as a small macro cell which are used to provide the network coverage in cases where the footprint of a macrocell is not necessary.

Pico cells provide the coverage and capacity in some areas inside the macro cell. Pico cells are lower-power station than macrocells with transmit power from 23 dBm to 30 dBm. They are deployed indoor or outdoor serving a few tens of users within a range of 300 meter or less. Pico cells are prefer to provide the public area such as transport station, shopping center.

\begin{figure}
  \centering
  \includegraphics[scale=0.6]{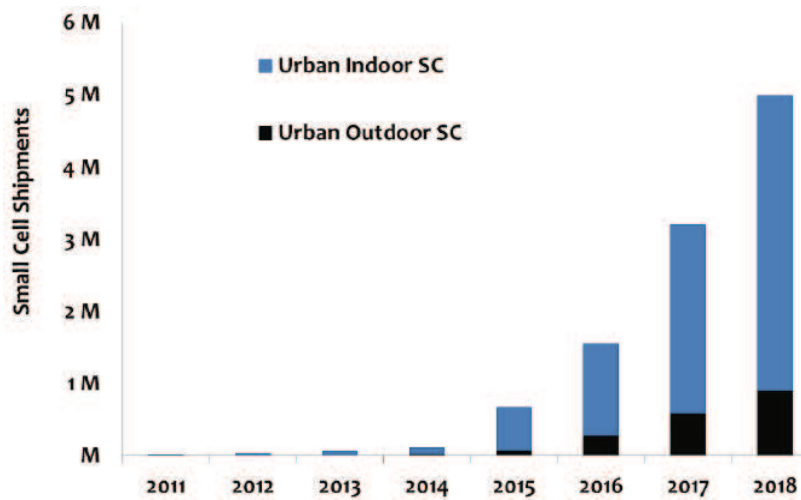}\\
  \caption{Smallcell Shipments~\cite{Market2014}}\label{lb-IndoorData}
\end{figure}

Femto cells are designed for offering network services inside apartment and office, and the femto-cell are prefer to work at Closed-Access Mode in which only allows the registered users authorizing by the owner. Since the low quality of received signal caused by the penetration losses through walls will degrade the performance of indoor date access, the deployment of small cell inside house can enhance the link capacity by reducing the transmission distance. The future mobile data traffic is forecasted that data access of smart devices will majorly generate indoor as shown in~Fig.\ref{lb-IndoorData}. Femto-cells are denoted as Home evolved NodeBs (HeNBs) placed inside house are low-cost, low-power consumption, short-range, plug-and-play base stations. Femto-cells are represented for the small cells in near future.

Thus, the advantage of the deployment of small cells is to the decrease in the transmission range between the sender and the receiver, which in turn enables the mobile subscriber to use the network services with high data speed at anywhere.

\begin{table}
\caption{Smallcell Nodes} 
\centering 
\begin{tabular}{c c c} 
\hline 
Node types & Transmit power & Coverage \\ 
\hline 
Macrocell                   & 43-46 dBm & few Km \\ %
Microcell                   & 23-33 dBm & $\leq$ 500 m \\ %
Picocell                    & 23-30 dBm & $\leq$ 300 m \\ %
Femtocell                   & $\leq$ 23 dBm & $\leq$ 50 m\\ %
\hline 
\end{tabular}
\label{Smallcell} 
\end{table}

\begin{figure}
  \centering
  \includegraphics[scale=0.6]{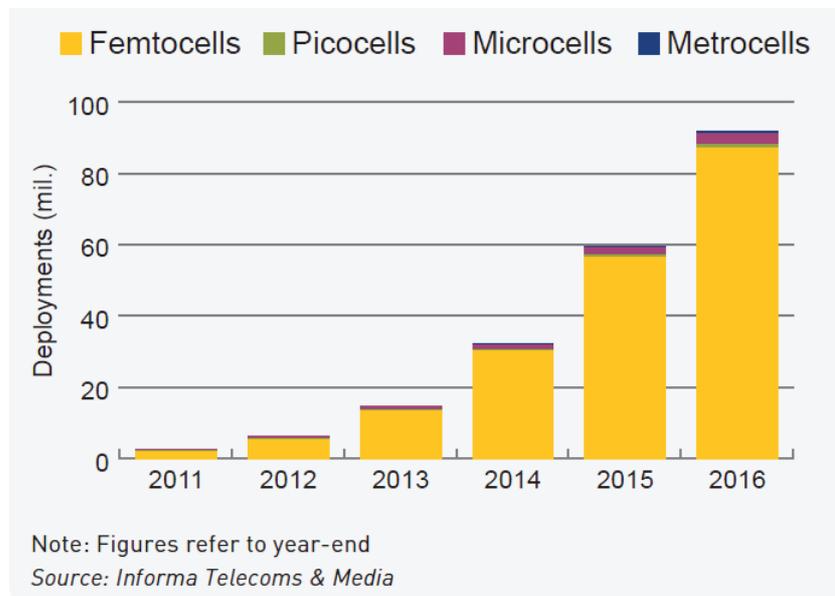}\\
  \caption{Global small-cell deployment forecasts, 2011-2016~\cite{Smallcell2012}}\label{lb-MarketShare}
\end{figure}

The specification of small cell node is summarized in Table.~\ref{Smallcell}. Small cell market forecasts to hit \$2.7 billion by 2017 according to Informa Telecoms \& Media in~\cite{Smallcell2012}. As expected, in South Korea, SK Telecom has deployed close to 40,000 small cells and is the leading adopter of 4G public access femto cells to date. As shown in~Fig.\ref{lb-MarketShare}, the portion of femto cells takes a large position as compared with the other. Femto cells are prefer to deploy in office and apartment in order to increase the  mobile services capacity and quality inside your apartment and office. Femto cells also expected as easy-deployed and convenient devices to the end users. Therefore, the more research focus on femto cells can prepare a good solution for  mobile networks in near future.

Small cells can be used to provide indoor and outdoor wireless services and recover cell-edge user performance and offload the macro cells. The network operators use small cells to extend the network coverage and increase the network capacity. In this paper, we consider the term of "small cells" referred as "femto cells" because femto cells are expected as emerging technologies that can meet the requirement of mobile data traffic over the next ten to fifteen years. In such networks, macro cells and femto cells share the radio frequency spectrum with each other. The re-use of resource can help increase network capacity and reach the peak data rate. Femto cells are prefer to work at Closed-Access Mode that allow a limited number of users to connect. Hence, this configuration creates the coverage hole inside the macro cells ($"black hole"$), the macro users located within the transmission range of HeNBs cannot be served by that HeNBs even experience high interference. For example, the quality of downlink signal from serving MeNBs to macro user can be affected by the downlink signal from nearby HeNBs. The deployment of small cells poses a such interference challenge, some efforts have been studied in order to limit this interference and manage the radio frequency efficiently that will be described in Section~\ref{Related}.

In this paper, we are interested in solving inter-cell interference (ICI) in macro femto network environment. To cope with ICI problem, enhanced inter-cell interference coordination (eICIC) techniques have been proposed in Release 10~\cite{R1-104256, R1-104968, R1-104661}. eICIC solutions include some techniques such as time-domain technique, frequency-domain technique, and power control technique. Almost Blank Sub-Frame (ABSF) is one of time-domain technique in which the interfering cell will stop using some sub-frames in order to reduce ICI.

ABSF technique is simple but efficient solution, the number of sub-frames is carefully chosen that can reach the optimal network performance. Unlike the previous work, in this paper, we consider the number of ABSF based on the quality of service (QoS) of the macro users. We also propose for setting the muted sub-frames efficiently based on cooperation between the mutual interfering HeNBs.


\section{Related Work}
\label{Related}
The research on eICIC technique has been studied in recent years for both macro-pico and macro-femtocell network~\cite{David2011a, MIKI2012}. In~\cite{Pedersen12}, the authors explain the benefits and characteristics of the enhanced inter-cell interference coordination (eICIC) technique. In macro-pico cell network, the range extension (RE) and time domain eICIC are used in pico-cell side, then the performance results are shown the recommended settings of the RE offset and the muting ratio in different scenarios in~\cite{Wang2012, Madan2010}. In~\cite{Jiang2012}, the authors propose a resource allocation scheme for eICIC in HetNets. The result of this paper shows the performance of using ABSF and UE partition scheme in which a fixed ABSF is set under each UE partitioning, and new ABSF pattern is selected for next time for all eNBs.

In macro-femto cell network scenario, in~\cite{Kamel2012} the authors also provide the performance of eICIC technique in HetNets with fixed muted rate. In~\cite{Kamel2013}, the authors propose an ABSF offset and resource partition, in which the number of ABSF is derived depends on the number of victim macro user and total macro user. Another paper finding the optimal number of ABSF is presented in~\cite{Lembo2013}. The result shows that in most case the number of muted sub-frames depends on the number of victim macro user, the number of normal macro user, and the femtocell user, and the muted rate is set globally for all eNBs. In~\cite{Cierny13}, the authors derive the number of ABSF in both macro-pico and macro-femto scenario for HetNets. In this paper, the HetNet scenario is modeled using stochastis geometry and required number of ABSFs is formulated based on base station placement statistics and user throughput requirement. The number of ABSF is set globally for all eNBs, it seems to be not work well in macro-femto cell network environment.

\section{Contribution}
\label{Contribution}

Resource allocation task is to distribute the radio resources among users with a limited transmit power, a specific time slot, and an exact frequency. The resource allocation is well organized, all users can have a good portion of resource to use that leads to reach a better performance of overall networks. In this paper, our goal is to develop an efficient ABSF framework in order to manage the interference downlink for HetNets~\citep{Capozzi2013}. To do that, we develop an algorithm based on the enhanced inter-cell interference (eICIC) technique in time-domain in order to mitigate the interference and enhance the overall network performance. The proposed algorithm includes two mechanisms: the first one is to select a dynamically optimal number of ABSF based on the quality of service of macro user, each HeNB works at different ABSF mode. The second one is to group HeNBs into a coalition in order to set the muted sub-frames among mutual interfering HeNBs efficiently.

The rest of this paper is organized as follows. Chapter~\ref{Chapter2} describes the system model and problem. Chaper~\ref{Chapter3} introduces the proposed algorithm for setting the number of ABSF required for MUEs and HeNBs and a framework for cooperation among interfering HeNBs. In Chaper~\ref{Chapter4}, we show the simulation results to demonstrate the performance of our proposal. We conclude the paper in Chaper~\ref{Chapter5}.


\chapter{System Model and Problem} 

\label{Chapter2} 

\lhead{Chapter 2. \emph{System Model and Problem}} 


\section{System Model}
\label{lb-SystemModel}
We consider the downlink (DL) of Heterogeneous Cellular Networks (HetNets) as depicted in Fig.~\ref{lb-Sys}. The macro base station MeNB is located at the center of each cell and the transmission of MeNB is fixed at 46 dBm. The macro users MUEs are randomly inserted inside the coverage of MeNB. In each apartment, one femtocell base station HeNB is located at the center of apartment (it can be deployed anywhere inside apartment), we assume that each HeNB serves only one femtocell user FUE. The maximum transmission power of HeNB is 20 dBm.

\begin{figure}
  \centering
  \caption{System Model}\label{lb-Sys}
\end{figure}

We assume that HeNBs work at Closed Access Mode, HeNBs do not allow MUEs to connect. If a MUE serving by the MeNB is located nearby a HeNB, it experiences high interference from nearby HeNB, then the desired signal from MeNB to MUE becomes weak. From now on, we denote the MUE that affected by nearby HeNB to be a victim macro user VMUE, and that HeNB is called a interfering HeNB or a aggressor HeNB.

The X2 interface is assumed to connect among the HeNBs using inter-cell inter-coordination (ICIC) in order to mitigate the ICI for MUEs. A basic ICIC technique involves resource coordination amongst interfering base-stations, where an interfering base-station gives up use of some resources in order to enable control and data transmissions to the victim user terminal. The MeNB and the HeNBs can exchange information via backhaul network, and the communication between MeNB and MUEs performs via OTA.

\section{Problem Definition}
\label{lb-Prob}
In this paper, we consider the heterogeneous network, which consists of macrocell network layer and femtocell network layer. When a macro user MUE is close to a femtocel access point HeNB, called victim MUE, the received signal from the macro base station (MeNB) to the MUE may interfere with the received signal from the HeNB. Hence, the victim MUE may not successfully decode and receive the signal from its serving macro base station MeNB. The 3GPP release 10 has proposed the eICIC techniques to improve the downlink capacity~\cite{R1-104968}. Almost Blank SubFrame (ABSF) is one of the time-domain technique of the eICIC, in which the HeNB will stop using some subframes to enable the victim MUE to exchange the data with its serving base station continuously until the downlink interference is diminished. The eICIC techniques will maintain the performance of the macro user in the present of femtocells. However, when the HeNB stops using the resources, the femtocell users belong to the HeNB can not use the downlink from the HeNB that leads to degrade the femtocell performance.

The blanking rate at HeNB helps improve the macro user performance, but affects the femto user performance degradation. The 3GPP has proposed the eICIC ABSF mechanism to enhance the network performance with a fixed blanking rate~\cite{R1-104256}, the performance of the proposal is evaluated in~\cite{Kamel2012}. In~\cite{Cierny13, Lembo2013, Kamel2013}, the authors presented an optimal ABSFs but fixed for all HeNBs. Since, the HeNB is located at difference place in the network coverage, it has an unique characteristics as illustrated in~Fig.\ref{lb-Sys}. For example, the present of HeNB-0 does not influence any macro user MUE; in contrast, others HeNB-1, HeNB-2, HeNB-3 create a downlink interference to surrounded macro users MUEs. Hence, the ABSFs for each HeNB should be optimal and difference in order to utilize the network resource effectively.

In this paper, we propose a downlink packet scheduling for Macro/ Femto-cell network to maximum the femto-cell throughput while maintaining the macro performance. We propose an approach to dynamically select the optimal ABSF pattern based on the Quality of Service (QoS) requirement of victim MUEs in each aggressor HeNB.
\section{Path Loss Model}
\label{Loss-model}
In this paper, we assume the path loss model according to the urban deployment scenario~\cite{R4-092042}. The path loss is modeled at different types of links depending on the position and type of users. We consider two kinds of links: the downlink between serving MeNB and MUE and the downlink between interfering HeNB and MUE.

The path loss between the serving MeNB and MUE can be expressed in~Table~\ref{Tab-PathLoss1}, where $D$ is the distance between MeNB and MUE in meter, and $L_{ow}$ is the penetration loss of an outdoor wall, which is 10 dB or 20 dB.

\begin{table}
\caption{Path loss between serving MeNB and MUE} 
\centering 
\begin{tabular}{l l} 
\hline 
Position of MUE & Path Loss (dB) \\ 
\hline 
MUE is outside of a apartment                  & $L(dB) = 15.3 + 37.6 \log_{10}D$ \\
MUE is inside of a apartment                   & $L(dB) = 15.3 + 37.6 \log_{10}D + L_{ow}$ \\ %
\hline 
\end{tabular}
\label{Tab-PathLoss1} 
\end{table}

The path loss between the interfering HeNB and MUE can be expressed in~Table~\ref{Tab-PathLoss2}, where $d$ is the distance between HeNB and MUE in meter.

\begin{table}
\caption{Path loss between interfering MeNB and MUE} 
\centering 
\begin{tabular}{l l} 
\hline 
Position of MUE & Path Loss (dB) \\ 
\hline 
MUE is inside or outside of a apartment                  & $L(dB) = 127 + 30 \log_{10} (d/1000)$ \\
\hline 
\end{tabular}
\label{Tab-PathLoss2} 
\end{table}

After calculating $L$, the shadowing model is considered, all links will take into account for shadowing model by adding log-normally distributed shadowing with standard deviation of 10 dB or 8 dB for links between MeNB and MUE, HeNB and MUE, respectively.

\section{Interference Model}
\label{IF-model}
Due to the deployment of small networks, the HetNets now consist of two-layers, the first layer is the macro cell network, while the second layer is the small cell network. In order to use the spectrum resource effectively and achieve high network capacity, the macro cell and small cell network share the same frequency bands. This situation brings a problem of interference management. The interference can be divided into two types: co-layer interference and cross-layer interference in~\cite{Zahir2013}.

The co-layer interference occurs when the network devices interfere each other in the same layer. For example, in small cell layer, the deployment of femtocell is random and mass, as they are close to each other in apartments. There is a possible way of power leaks through windows, doors, and balconies that leads to the interference among the neighboring femtocells. The second type of interference is cross-layer interference caused by the network devices that belong to different layer of network. The femtocell base stations are usually deployed by the end users their home, apartment, or office. The owners prefer the femtocell base stations work at the closed-access mode that only allows the registered-subscriber to access the service offered by that femtocells base stations. At the same time, the femtocell base stations create the coverage hole inside the coverage of macro cell network. Hence, the femtocell base stations can cause interference to the downlink of macro user nearby. In this paper, our interest is to cope with the cross-layer interference, we address the problem of downlink interference to improve the overall network performance.

In HetNets, each cell consists of a macro base station (MeNB) serving $\mathcal{M}$ macro users (MUEs). $\mathcal{F}$ small cell base stations (HeNBs) are randomly inserted inside the coverage of MeNB, each HeNB $F_{f}$ works at closed access mode that only allows one small cell user (FUE) to connect. To reach the peak data rate, all network nodes will use all bandwidths. Then the downlink signal from MeNB to MUEs interferes with the downlink signal from nearby HeNB to that MUEs as depicted in Fig.~\ref{lb-Sys}.

The signal-to-interference-and-noise ratio (SINR) $\gamma_{m}$ at link $\mathcal{L}_{m}$ between the MeNB and the macro user MUE $m$ is defined as
 \begin{eqnarray} \label{eq:sinr-m}
  \gamma_{m} &=& \displaystyle \frac{G(M, m) P(m)}{\Sigma _{f \in \mathcal{F}} P(F_{f}) G(F_{f}, m) + \sigma_{m}} \\
             &=& \displaystyle \frac{G(M, m) P(m)}{\eta_{m}} \nonumber
 \end{eqnarray}
where $\sigma_{m}$ is the thermal noise at macro user $m$. Let denote $P(m)$ and $P(F_{f})$ to be the transmission power of MeNB and HeNB, respectively, $G(M, m)$ is the path gain between the macro base station MeNB and macro user $m$, and $\eta_{m}$ is the sum of interference and noise at macro user $m$.

Let denote $\mathcal{L}$ to be the set of links from the MeNB to their serving MUEs, $\mathcal{L} = (\mathcal{L}_{1}, ..., \mathcal{L_{\mathcal{M}}})$. To ensure the quality of signal from MeNB to each user MUE, each link $m$ has a minimum requirement in terms of SINR, i.e $\gamma_{m} \geq \gamma_{0}$. The minimum SINR requirements can be rewritten in matrix form as
  \begin{equation} \label{eq:sirn-matrix}
  \mathbf{F} \mathbf{P}_{\mathcal{F}} \leq  \mathbf{P}_{\mathcal{M}} - \mathbf{b}, \nonumber
 \end{equation}
 where $\mathbf{P}_{\mathcal{M}} = (P(1), ..., P(\mathcal{M}))^T$, $\mathbf{P}_{\mathcal{F}} = (P(F_{f}), ..., P(F_{\mathcal{F}}))^T$, $\mathbf{b} = (b(1), ..., b(\mathcal{M}))^T$ such that $b(m) = \displaystyle \frac{\gamma_{0} \sigma_{m}}{G(M, m)}$, and $\mathbf{F}$ is a non-square matrix with $\mathcal{M} \times \mathcal{F}$ elements that can be defined as
\begin{equation} \label{eq:sirn-matrixF}
\mathbf{F}(m,f) = \displaystyle \frac{ G(F_{f}, m) \gamma_{0}}{G(M, m)}.
\end{equation}

If the value of element $\mathbf{F}(m,f)$ is non-zero, this means that the HeNB $F_{f}$ creates a downlink interference to the corresponding MUE $m$. The MeNB can determine the interfering HeNBs to the victim MUEs, and then ask the interfering HeNBs to work at the ABSF mode. The ABSF mode will be described later.

\section{Overview of eICIC techniques}
\label{lb-eICIC}
As mentioned above, the eICIC techniques have been discussed in RAN1 focusing on the macro-femto deployment. Three candidate solution for eICIC have been proposed. The first solution is power control in which the HeNB adjust its transmission power to avoid interference to other. The second frequency-domain solution uses orthogonal bandwidth for control signalling an common information. The last one is time-domain solution by stopping use some sub-frames the interfering node can reduce its downlink interference to the users. A summary of description of candidate eICIC is discussed in~\cite{R1-104968}.

In power control approach, the power allocation can be determined based on some strategies such as the strongest receiving power of MeNB at the Femto, the pathloss between Femto base station and macro user MUE, the objective SINR of HUE, and the objective SINR of MUE. In frequency-domain solution, reduced bandwidth for control channels and physical signal, such that the control channels and physical signals can be totally orthogonal to those in another layer.

\begin{figure}
  \centering
  \includegraphics[scale=0.36]{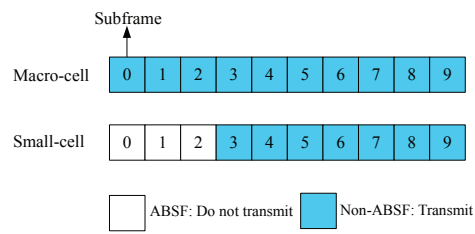}\\
  \caption{The Almost Blank Sub-Frame}\label{lb-absf}
\end{figure}

In time-domain solution, some sub-frames called Almost Blank Sub-Frames (ABSFs) are muted at the interfering HeNBs, then the MeNB can schedule its downlink to the victim MUE in order to recover the signal quality of its user as shown in Fig.~\ref{lb-absf}. The use of ABSF can help MUE recover its performance due to the strong downlink interference of HeNBs, in another side, the performance of the FUE serving by the interfering HeNB is reduced. Hence, the selection of number of sub-frames and which and when HeNB should start to mute are crucial in eICIC ABSF approach.

\section{Overview of LTE Frame Structure}
LTE~\cite{Ghosh2010} employs OFDMA to allocate the radio resources for downlink transmission. Users are allocated a specific number of sub-carriers for a predetermined amount of time called physical resource blocks (PRBs) in the LTE specifications. In order to explain the OFDMA, the LTE frame structure is studied.

As shown in Fig.~\ref{lb-DL-RB}, each radio frame is divided into 10 sub-frames with 1 millisecond length of time. Each sub-frame is future divided into two time slots, each of 0.5 millisecond duration. Slots consist of either 6 or 7 OFDMA symbols, depending on whether the normal of extended cyclic prefix is used.
\begin{figure}
  \centering
  \caption{The structure of the downlink resource grid~\cite{Ghosh2010}}\label{lb-DL-RB}
\end{figure}

The total number of available sub-carriers varies according to the overall transmission bandwidth system used. For example, the total bandwidth is varying from 1.25 MHz to 20 MHz, then the number of available resource block is increasing from 6 to 100 resource blocks. Each resource block (RB) is defined as consisting of 12 consecutive sub-carriers for one time slot in duration called resource element with 15 kHz of bandwidth. Each PRB is smallest element of radio resource, each user is assigned to a specific PRB for their uplink or downlink transmission.

Let $N_S$, $N_{RB}$ denote the number of sub-frames for each radio frame and the number of resource blocks in LTE downlink structure as displayed in~Fig.\ref{lb-DL-RB}, respectively.

\chapter{Proposed Algorithm} 

\label{Chapter3} 

\lhead{Chapter 3. \emph{Proposed Algorithm}} 


In this section, we propose an algorithm to select dynamically optimal ABSFs for each HeNB, and to group the interfering HeNB into coalition in order to decide which and where subframes should be muted by the interfering HeNB in each coalition. The muted rate at each HeNB is selected as minimum as possible to enhance the performance of the femtocell users while satisfies the QoS of the macro users. Our proposed algorithm firstly determine the required muted rate for each VMUE via ABSF selection mechanism. By applying the second mechanism called Interfering HeNBs Coaliation, the aggressor HeNBs are grouped into coalition, each aggressor HeNB with a different muted rate and a time slot has to stop their transmission on time.

\section{ABSF Selection}
\label{ABSF}
When the HeNB is marked as an aggressor HeNB, it will stop using some subframes to limit its interference to the victim MUE. Then the SINR $\gamma_{m}$ at link $\mathcal{L}_{m}$ between the MeNB and the macro user MUE $m$~(\ref{eq:sinr-m}) can be rewritten as
 \begin{eqnarray} \label{eq:sinr-m-alpha}
  \gamma_{m} &=& \displaystyle \frac{G(M, m) P(m)}{\Sigma _{f \in \mathcal{F}} P(F_{f}) G(F_{f}, m)(1 - \alpha_{m}) + \sigma_{m}}
 \end{eqnarray}
 where $\alpha_{m}$ is the muted rate required for macro user $m$ during one radio frame. Our objective is to find the muted rate in order to satisfy the required minimum SINR of the victim MUE.

\begin{equation}\label{eq:5}
\begin{aligned}
& {\text{minimize}}
& & \sum \alpha_{m} \nonumber \\
& \text{subject to}
& &  \mathbf{A} \alpha_{m} \succeq \mathbf{B},\nonumber \\
& & & 0 \leq \alpha_{m} \leq 1,
\end{aligned}
\end{equation}
where $\mathbf{A} = \mathcal{F} \mathbf{P}_{\mathcal{F}}$, and $\mathbf{B} = \mathbf{P}_{\mathcal{F}} - \mathbf{P}_{\mathcal{M}} + \mathbf{b}$. The operation $\succeq$ here is used to said that element-wise operation for each $\alpha_{m}$.

Since $\mathbf{A}$ is a non-square matrix, a solution to problem $\mathbf{A} \alpha_{m} = \mathbf{B}$ that minimizing $\alpha_{m}$ is
\begin{equation}\label{eq:solution}
 \alpha_{m}^{\ast} =  \mathbf{A}^{T} (\mathbf{A} \mathbf{A}^{T})^{-1} \mathbf{B}.
\end{equation}

The aggressor HeNB may be asked to work with different muted rates for different macro users. Hence, to satisfy all victim macro users MUEs belonging to the aggressor HeNB $F_{f}$, the muted rate for the HeNB $F_{f}$ is $\alpha_{F_{f}}$, $\alpha_{F_{f}} = \max (\alpha_{m}^{\ast}), m \in \mathcal{H}_{f}$, $\mathcal{H}_{f}$ is a set of victim macro users VMUEs that affected by HeNB $F_{f}$. For instant, each macro user asks for different muted rate and then the muted rate for the HeNB can be calculated based on the required rate for the victim MUEs as shown in Table~\ref{alpha-m} and~\ref{alpha-f}.

\begin{figure}
  \centering
  \caption{A coalition of aggressor HeNBs}\label{lb-cluster}
\end{figure}

\begin{table}
\caption{Muted Rate for Macro User - MUE} 
\centering
\begin{tabular}{| c | c | c |} 
\hline 
MUE & $\alpha_{m}$ & HeNB \\ 
\hline 
1                 & $\displaystyle \frac{1}{10}$ & 1\\ 
2                 & $\displaystyle \frac{2}{10}$ & 1, 2, and 3\\ 
3                 & $\displaystyle \frac{2}{10}$ & 2 and 3\\ 
4                 & $\displaystyle \frac{2}{10}$ & 1 and 2\\ 
5                 & $\displaystyle \frac{2}{10}$ & 2 and 3\\ 
\hline 
\end{tabular}
\label{alpha-m} 
\end{table}

\begin{table}
\caption{Muted Rate for Femtocell Base Stattion - HeNB} 
\centering
\begin{tabular}{| c | c |} 
\hline 
HeNB & $\alpha_{F_{f}}$  \\ 
\hline 
1                 & $\displaystyle \frac{2}{10}$ \\ 
2                 & $\displaystyle \frac{2}{10}$ \\ 
3                 & $\displaystyle \frac{2}{10}$ \\ 
\hline 
\end{tabular}
\label{alpha-f} 
\end{table}
\section{Interfering HeNBs Coaliation}
\label{Coaliation}
In urban dense small-cell network, the number of macro users and small-cell base stations becomes extremely larger and uncontrollable. In order to reduce interference in this scenario, the neighbor small-cell base stations should form into a group or a coalition, where two or more small-cell base stations create a downlink interference to the same macro users. The interfering small-cell base stations in that coalition should be muted at the same sub-frames in order to reduce the downlink interference efficiently. An example is illustrated in~Fig.\ref{lb-cluster}, the VMUE-4 affected by two neighbor HeNBs 1 and 2, if the HeNB-1 mutes at sub-frame $s_{1}$ and the HeNB-2 mutes at sub-frame $s_{2}$, then during the sub-frames $s_{1}$ and $s_{2}$ the victim MUE still affected by either HeNB-2 or HeNB-1. Hence, a VMUE is be affected by multiple HeNBs in the coalition, the group of that HeNBs has to stop their data transmission at the same sub-frames in order to reduce the interference efficiently and increase overall Quality of Experience (QoE).

The deployment of small-cell base stations is unplanned and massive, the network operator will face how to manage the radio resources and the interference between the macro base stations and the small-cell base stations. Hence, the small-cells should be self-organization and self-configuration to form cooperative groups via X2 interface.

To do so, firstly the number of VMUEs affected by each HeNB will be listed. Then, based on the VMUEs list, the aggressor HeNBs can cooperate to form coalition. To collect the VMUEs that affected by an aggressor HeNB, the process is performed as follows. If the SINR of the macro user $m$, $\gamma_{m}$, is less than the threshold SINR, $\gamma_{0}$, then the macro user $m$ is marked as a VMUE. Each MUE can determine its status and report back to the serving MeNB by computing the matrix $\mathbf{F}(m,f)$, each value in each row (MUE) maps to a corresponding column (HeNBs). If the value is non-zero, the corresponding HeNB will be marked as an aggressor HeNBs. We now obtain the list of aggressor HeNBs that affects each VMUE. Any aggressor HeNB affects VMUE, it will insert the VMUE into its set of affected VMUEs. The operation of the VMUEs collecting process is listed in Algorithm~\ref{A-I}. The whole process is performed by the MeNB, then the MeNB will send the set of VMUEs of each aggressor HeNB to corresponding aggressor HeNBs.  At the beginning, the first aggressor HeNB checks its name in a set of aggressor HeNBs of each VMUE $v_{n}$, if it appears, the aggressor HeNB inserts the VMUE $v_{n}$ into its affected list. The loop will perform until the last aggressor HeNB.

\begin{algorithm}
\caption{The VMUEs Collecting Algorithm}
\begin{algorithmic}
\STATE $\mathcal{V}$: the set of all victim MUEs $\mathcal{V} = \{v_{1}, v_{2}, ...,v_{\mathcal{N}}\}$
\STATE $\mathcal{A}_{n}$: the set of aggressor HeNBs affected VMUE $v_{n}$
\STATE Find the set of VMUEs affected by an aggressor HeNB $F_{f}$: $\mathcal{H}_{f}$
\FOR{$f = 1$ to $\mathcal{F}$}
\STATE $\mathcal{H}_{f} = \O$
\FOR{$n = 1$ to $\mathcal{N}$}
\IF{$\{F_{f} \} \cap \mathcal{A}_{n} \neq \O$}
\STATE $\mathcal{H}_{f} = \mathcal{H}_{f} \cup \{v_{n}\}$
\ENDIF
\ENDFOR
\ENDFOR
\end{algorithmic}
\end{algorithm}\label{A-I}

To group the mutual interfering HeNBs coalition, the aggressor HeNBs cooperate to find which aggressor HeNB is grouped into a coalition as illustrated in Algorithm~\ref{A-II}. The main idea of Algorithm~\ref{A-II} is that neighbor HeNBs exchange their information via X2 interface, if two or more aggressor HeNBs have same VMUEs, they are grouped into a coalition, each aggressor HeNBs can only belong to one coalition. If two coalitions have same HeNB, they will joint together as one coalition.

\begin{algorithm}
\caption{The Mutual Interfering HeNBs Grouping Algorithm}
\begin{algorithmic}
\STATE $\mathcal{H}_{f}$: the set of VMUEs affected by an aggressor HeNB $F_{f}$
\STATE ${flag}_{f}$: a flag associated with $\mathcal{H}_{f}$. $\mathcal{H}_{f}$ is grouped, ${flag}_{f} = 1$, if not ${flag}_{f} = 0$
\STATE $\mathcal{C}_{c}$: the set of VMUEs affected by a group of mutual interfering HeNBs
\STATE $\mathcal{G}_{c}$: the set of mutual interfering HeNBs
\FOR{$c = 1$, $f = 1$ to $\mathcal{F}$}
\IF{${flag}_{f} = 0$}
\STATE $\mathcal{C}_{c} = \mathcal{H}_{f}$, $\mathcal{G}_{c} = \{F_{f}\}$, ${flag}_{1} = 1$
\FOR{$f = 1$ to $\mathcal{F}$}
\IF{$\mathcal{C}_{c} \cap \mathcal{H}_{f} \neq \O$ and ${flag}_{f} = 0$}
\STATE $\mathcal{C}_{c} = \mathcal{C}_{c} \cup \mathcal{H}_{f}$, $\mathcal{G}_{c} = \mathcal{G}_{c} \cup \{F_{f}\}$
\ENDIF
\ENDFOR
\STATE $c = c + 1$
\ENDIF
\ENDFOR
\end{algorithmic}
\end{algorithm}\label{A-II}

After applying Algorithm~\ref{A-I} and~\ref{A-II}, the smallcell networks are partitioned into coalitions. 
The details of our proposed scheme is shown in~Fig.\ref{lb-Algorithm}. Firstly, the VMUEs are marked via the channel-quality indicator (CQI) if the SINR of VMUEs is low, then each  VMUE $v_{n}$ estimates the required muted rate $\alpha_{m}$ for itself.  Each VMUE $v_{n}$ reports the set of aggressor HeNB $\mathcal{A}_{n}$ affecting its downlink to the MeNB. The MeNB performs the Algorithm~\ref{A-I} to collect the VMUEs set $\mathcal{H}_{f}$ that affected by the HeNB $F_{f}$. The list of VMUEs corresponding to an aggressor $F_{f}$ is sent to that HeNB. To group a mutual interfering coalition $\mathcal{G}_{c}$, the aggressor HeNBs cooperate via X2 interface by using the process of the Algorithm~\ref{A-II}.

\begin{figure}
  \centering
  \caption{The proposed algorithm}\label{lb-Algorithm}
\end{figure}


\chapter{Performance Evaluation} 

\label{Chapter4} 

\lhead{Chapter 4. \emph{Performance Evaluation}} 


\section{Performance of Proposed Scheme in Multi Users Scenario}
\label{lb-pf}
In this section, multi-users are considered including multi-macro users, multi-femtocell base station. The macro users MUEs are uniformly distributed inside the coverage of MeNB, and several HeNBs are located around the MUE. We assume that there are 0-4 HeNBs that are considered as the aggressors HeNBs for each MUE. To show the impact of location on the interference, we assume that each MUE is moving gradually away from the interfering HeNBs. Then, as expected the MUEs are free-interference user and get rid of the impact of the interfering HeNBs on their downlink signal.

We estimate analytically the dynamically optimal number of ABSF required for a macro user MUE in the multi users scenario. Since the SINR of MUE highly depends on the location between itself and MeNB or HeNB, after setting the scenario, the distance between MUE and HeNB is gradually increased that leads to the change in interference.

Our proposed algorithm is compared with the previous work with different muted rate. Let denote Optimal SINR, SINR - I, SINR - II, SINR - III be our proposal, and previous work with $\alpha = \left\{ \displaystyle \frac{1}{10}, \displaystyle \frac{2}{10}, \displaystyle \frac{3}{10}\right\}$, respectively.

The system scenario assumption and parameter are set up based on Monte-Carlo simulation~\cite{TR-36942}. The parameter settings are summarized in~Table~\ref{parameter}. The requirement of the SINR threshold is either -6 dB or -4 dB, in this paper we assume the threshold SINR of MUE, $\gamma _0$, is set to 0 dB.

\begin{table}
\caption{Parameter settings} 
\centering
\begin{tabular}{l l} 
\hline 
Parameter & Values \\ 
\hline 
System bandwidth                    & 10 MHz \\ 
Duplex Technique                    & FDD mode \\ 
Channel Model                       & Urban Macro-Femto Scenario Model \\ 
Frequency Reuse Scheme              & Reuse-1 \\ 
MeNB Range                          & 500~m\\ %
MeNB Transmission Power             &46~dBm\\ %
MeNB Antenna Gain                   & 14 dBi\\ %
Number of MUEs                      & 10-100 moving with random velocity\\
MUE Antenna Gain                    & 0 dBi\\ %
Number of HeNBs                     & 40-400 HeNB with transmit power of 20~dBm \\
HeNB Antenna Gain                   & 2.2 dBi\\ %
Thermal Noise                       & -174 dBm/Hz \\
Noise Figure                        & 9 dB  \\
Simulation Run Times                    & 2000 times run \\
\end{tabular}
\label{parameter} 
\end{table}

The SINR of all MUEs are collected and averaged for each step. Step here is defined as each time the distance between MUE and HeNBs is randomly gradual increased with the goal of decreasing the impact of interference. The number of steps is larger, this means that the distance between MUE and HeNB is larger. Hence, after a certain number of steps, the impact of interference on the signal is reduced approximately to zero.

\begin{figure}
  \centering
  \includegraphics[scale=0.7]{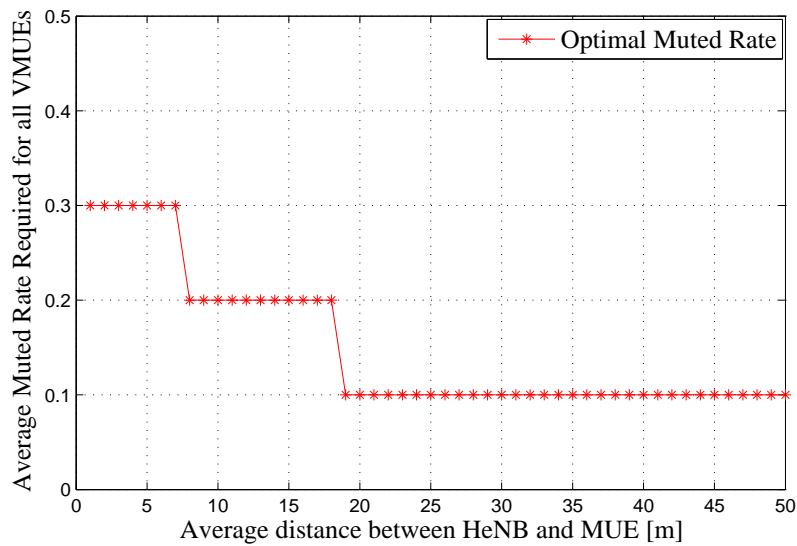}\\
  \caption{The muted rate required by the Victim MUEs}\label{S-MutedRate}
\end{figure}

We derive the muted rate required for the MUE as shown in~Fig.\ref{S-MutedRate}. The number of ABSF required for the MUE can be calculated by multiplying the muted rate and the number of sub-frames in each radio frame. The result show that the muted rate becomes zero when the MUE is far away HeNBs, the number of ABSF required is changing time by time. Hence, if ABSF pattern is set to a fixed value, the demand of the MUE may be higher or lower than the fixed value. A dynamically optimal muted rate is the best solution for eICIC technique in time domain.

\begin{figure}
  \centering
  \includegraphics[scale=0.7]{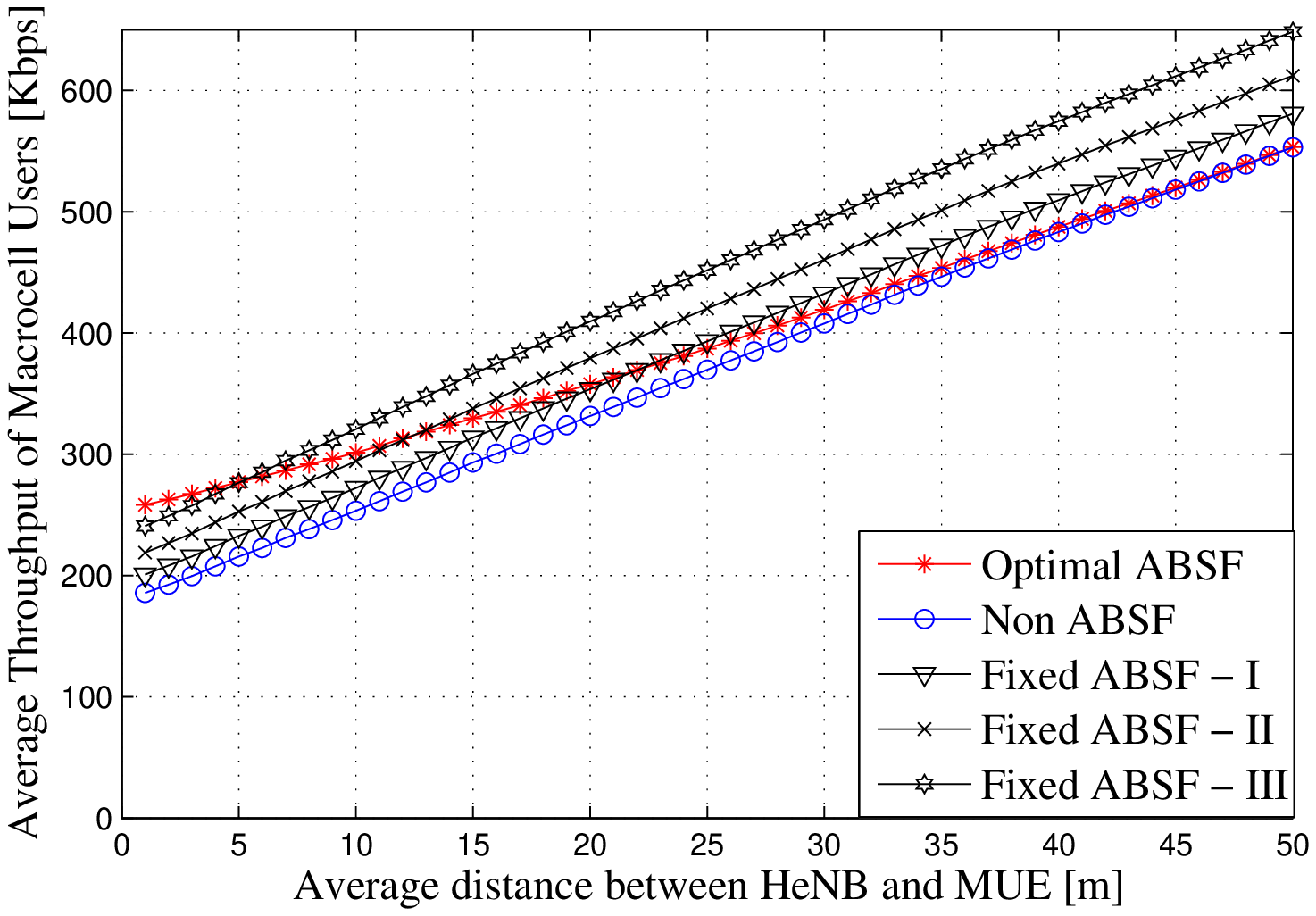}\\
  \caption{The optimal SINR of the MUE after using eICIC}\label{S-Optimal_SINR}
\end{figure}

In~Fig.\ref{S-Optimal_SINR}, the SINR of the MUE is reported as the lowest bound. As increasing the distance between the MUE and the HeNBs, the SINR of the MUE increases. Because the MUE is located far way the HeNBs, the impact of downlink interference from HeNBs is decreased, the desired signal from MeNB becomes good. When the distance between MUE and HeNBs is greater than a threshold distance, the impact of downlink interference on the desired signal of MUE can be tolerated. The optimal SINR of the MUE of our proposal and previous work are also shown in~Fig.\ref{S-Optimal_SINR} after apply the muted rate for all HeNBs in order to recover the performance of the MUE in the present of HeNBs. As can be seen in in~Fig.\ref{S-Optimal_SINR}, when ABSF mode is used, the performance of MUE is much better than without using ABSF mode.

\begin{figure}
  \centering
  \includegraphics[scale=0.66]{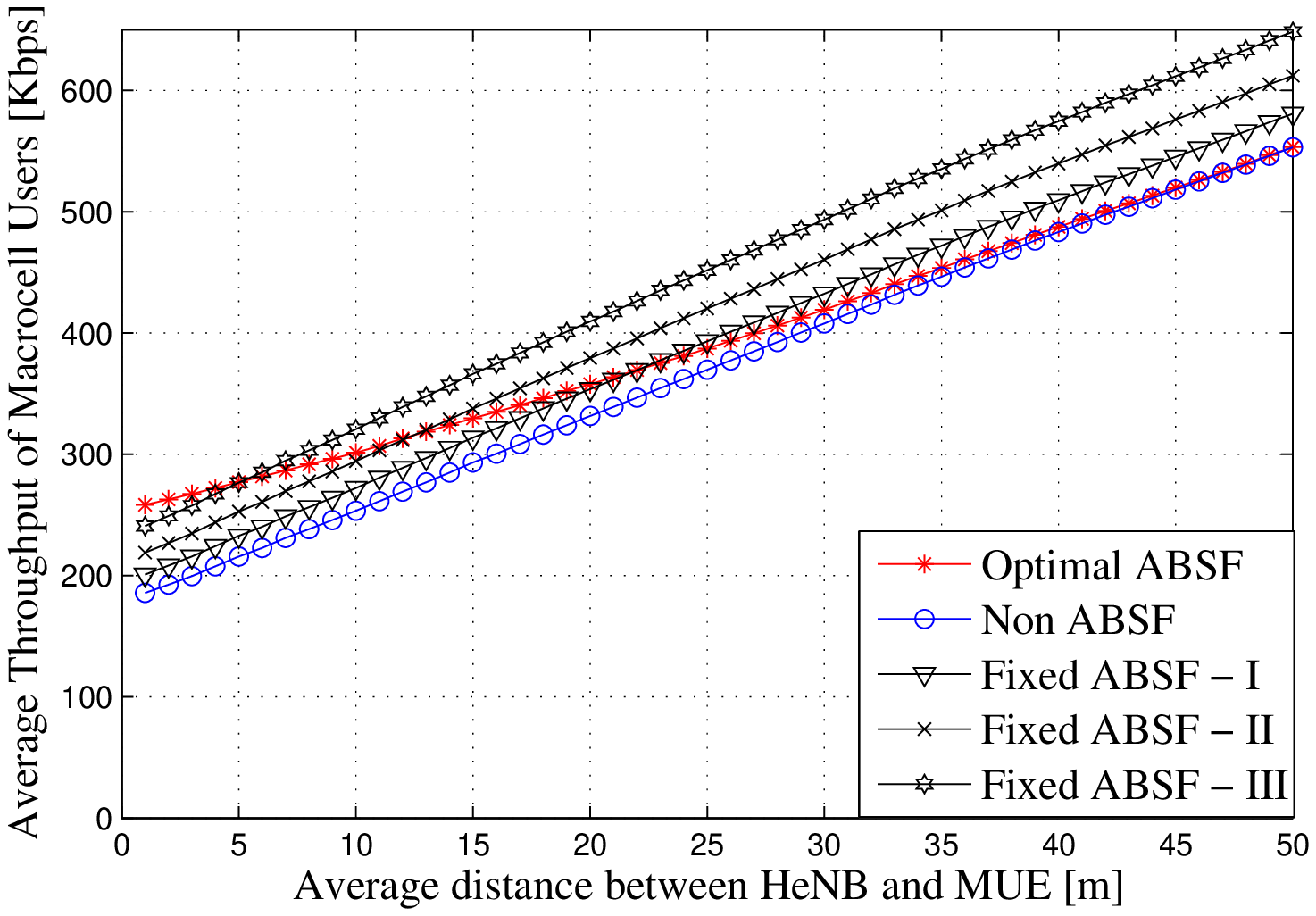}\\
  \caption{The achieved SINR of the MUE after using eICIC}\label{S-Achieved_SINR}
\end{figure}

The SINR gain of the MUE is shown in~Fig.\ref{S-Achieved_SINR} defined as the different between the SINR of MUE before and after applying the ABSF mode. Our proposal can adapt well adapt to the change of the impact of interference, while the ABSF pattern with fixed blanking rate may recover the performance of the MUE when the SINR of the MUE is low, but when the SINR of the MUE is high, the high blanking rate is harmful to the performance of the femtocells.

\begin{figure}
        \centering
        \includegraphics[scale=0.66]{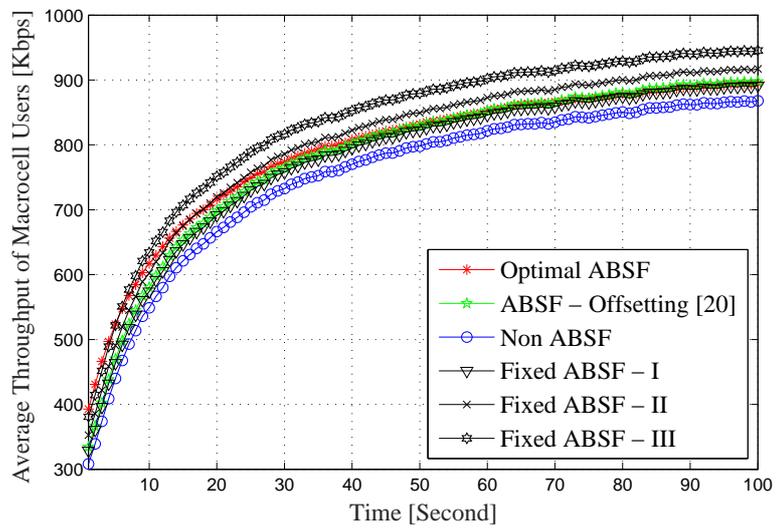}
        \caption{Average Macrocell Users Throughput [kbps] in case of muted rate of 0.3}
        \label{fig:Thr-Case3-Macro}
\end{figure}

\begin{figure}
        \centering
        \includegraphics[scale=0.66]{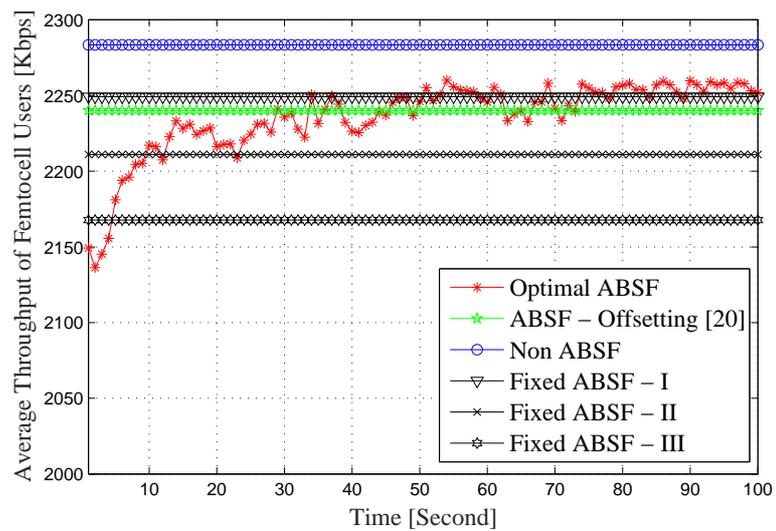}
        \caption{Average Femtocell Users Throughput [kbps] in case of muted rate of 0.3}
        \label{fig:Thr-Case3-Femto}
\end{figure}

We compare our proposed algorithm with previous work in terms of throughput. In Fig.~\ref{fig:Thr-Case3-Macro}, the throughput of macro users is reported. At the beginning steps, our proposed algorithm outperforms the previous work in th. As the number of steps increases, the performance of our proposed algorithm is lower than that of previous work. In contrast, in Fig.~\ref{fig:Thr-Case3-Femto}, at the beginning steps, the throughput performance of femtocell users of our proposed algorithm is less than that of previous work. Then the performance of femtocell users are recovered as number of steps increases. The reason is that at the beginning steps, when the MUEs locate near HeNBs, they require high muted rate at HeNBs, then the throughput of macro uses is good, but the throughput performance of femtocell users is poor. In the contrast, when the MUEs locate far away from the HeNBs, they require low muted rate, then the performance of the MUEs is smaller than that of previous work. In this time, the performance of femtocell users is good.

\begin{figure}
        \centering
      \includegraphics[scale=0.66]{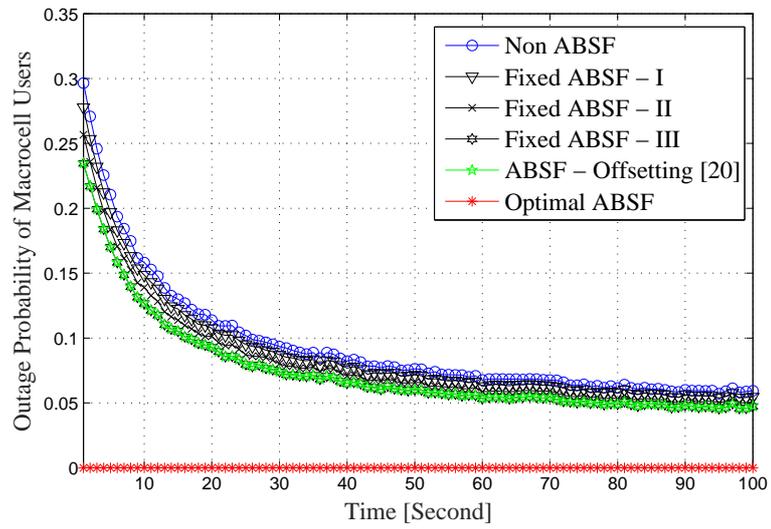}
      \caption{Outage Probability of Macro Users}
      \label{fig:Outage_M}
\end{figure}

Figure.~\ref{fig:Outage_M} shows the outage probability of macro users. The outage probability is defined as the ratio of number of unsatisfied macro users to total number of macro users. Our proposed algorithm is developed in order to satisfy the QoS of all macro users, there are no macro users that having their SINR less than the SINR threshold. In contrast, the previous work with fixed muted rate can not meet the QoS of macro users all cases even the high muted rate. Hence, our proposed algorithm outperforms the previous work.

\chapter{Conclusion and Future Work} 

\label{Chapter5} 

\lhead{Chapter 5. \emph{Conclusion}} 

\section{Conclusion}
\label{ccc}
This paper proposes a dynamically optimal ABSF eICIC framework in order to mitigate the impact of cross-layer interference in HetNets. Unlike previous work, the number of ABSF depends on the number of victim macro users, total number of macro users and femtocell users, and the number of ABSF is globally set for all HeNBs. In this paper, the number of ABSF is derived based on the QoS of each macro user MUE, and then based on the required muted rate for each MUE we can set a dynamically optimal blanking rate for each HeNB. Obviously, each HeNB owns a unique characteristic, then each HeNB has to decide its own muted rate as a local and distributed way. Due to the mutual interference among HeNBs, HeNB should cooperate to manage the resource allocation. We also propose a coalition algorithm to help HeNBs perform the ABSF framework efficiently. Via simulation, we show that our proposed algorithm outperforms the previous work.

\section{Publications}
\label{pubiii}
\cite{vu2014mobility} We propose a mobility-assisted on-demand routing algorithm for mobile ad hoc networks in the presence of location errors. Location awareness enables mobile nodes to predict their mobility and enhances routing performance by estimating link duration and selecting reliable routes. However, measured locations intrinsically include errors in measurement. Such errors degrade mobility prediction and have been ignored in previous work. To mitigate the impact of location errors on routing, we propose an on-demand routing algorithm taking into account location errors. To that end, we adopt the Kalman filter to estimate accurate locations and consider route confidence in discovering routes. Via simulations, we compare our algorithm and previous algorithms in various environments. Our proposed mobility prediction is robust to the location errors.

\cite{vu2014impact} In this paper, we analyze the impact of mobility prediction on ad-hoc on-demand routing algorithms in mobile ad-hoc networks. Location awareness enables mobile nodes to estimate link duration based on neighboring node mobility and choose the most reliable route. The estimated link duration also is used to adjust the hello interval in order to lessen the number of hello messages. We analyze the impact of on-demand routing algorithm with mobility prediction on the network performance in term of communication overhead when the largest path duration is chosen as the best route and the hello interval is adaptively adjusted according to mobility. Simulation results show a significant improvement of network performance. The total number of overhead messages is reduced by about 75$\%$ and 45$\%$ as compared with previous algorithm in low and high mobility environments, respectively.

\cite{vu2015cooperative} Vu, Trung Kien, Kwon, Sungoh, and Oh, Sangchul. "Cooperative Interference Mitigation Algorithm in Heterogeneous Networks." IEICE Transactions on Communications 98.11 (2015): p2238-2247.

\addtocontents{toc}{\vspace{2em}} 

\appendix 



\addtocontents{toc}{\vspace{2em}} 

\backmatter


\label{Bibliography}

\lhead{\emph{Bibliography}} 

\bibliographystyle{unsrtnat} 

\bibliography{Bibliography} 

\end{document}